# Development of a Mouse for Individuals without Upper Limbs using Arduino Technology

Alfonso Gunsha-Morales
*Facultad de Ciencias de la Ingeniería*
*Universidad Técnica Estatal de Quevedo*
Quevedo-Ecuador
0000-0002-1485-1522

Pedro Pardo
*Carrera de Telecomunicaciones*
*Instituto Superior Tecnológico Huaquillas*
Huaquillas-Ecuador
pardopedro46@gmail.com

Luis Enrique Chuquimarca Jimenez
*FACSISTEL*
*Universidad Peninsula de Santa Elena*
La libertad-Ecuador
0000-0003-3296-4309

David Herrera
*Carrera de Telecomunicaciones*
*Instituto Superior Tecnológico Huaquillas*
Huaquillas-Ecuador
jdherrera@isthuaquillas.edu.ec

**Abstract** – This project focuses on the design and construction of a prototype mouse based on the Arduino platform, intended for individuals without upper limbs to use computers more effectively. The prototype comprises a microcontroller responsible for processing signals from the MPU-6050 sensor, used as a reference for cursor position, and foot-operated buttons for right and left-click functions. Its design enables cursor control through head movements, providing users with an easy and intuitive way to interact with the computer's graphical interface. Feasibility testing was conducted through experimental trials, resulting in ideal accuracy and precision. These trials indicate that the device is viable for use in individuals without upper limbs.

## I. INTRODUCTION

The use of Information and Communication Technologies (ICT) is prevalent in personal, professional, and educational domains, offering various services to users. Common applications include information retrieval, new forms of communication, virtual learning, online banking, and e-commerce. Individuals with disabilities who face physical limitations hindering their access to ICT often experience a lack of digital skills, preventing them from acquiring knowledge, developing skills, and competing in the job market [1].

Many disabled individuals without upper limbs often refrain from using computers due to difficulties in using the peripherals designed for control. Market alternatives are either scarce or come at a high cost [2].

Hence, there is a pertinent need for the development of devices that cater to the specific needs of individuals with special abilities, enabling them to manipulate ICT and access the benefits offered by these technologies [3].

This is why the investigation into the implementation of an economically accessible mouse becomes relevant. Such a mouse would allow individuals without upper limbs to use their head and feet for navigating the graphical interface of an operating system and interacting with various programs in a way that is easy and interactive for the user.

Section 2 details the methodology applied in the development of this work, covering both hardware and software designs, along with system implementation and testing. Section 3 presents the results with detailed analysis, while Section 4 contains the conclusions derived from integrating these findings with the initial objectives of the project.

## II. METHODOLOGY

First, an analysis of the operation of a conventional mouse was conducted to transfer the same functions and adapt them to the needs of individuals without upper limbs [4] [5] [6]. The prototype is a wired mouse that allows cursor movement through head gestures and performs right and left-click functions using foot-operated pedals [7] [8] [9].

Subsequently, the prototype was manufactured by completing the final assembly of the electronic circuit on perforated micro-bakelite. Experimental tests were then conducted with the accelerometer of the MPU-6050 sensor incorporated into the GY-521 module [10] [11], connected to the Arduino Leonardo platform along with two pedal-operated buttons, implementing the configuration of 2 pull-down resistors [12] (see Figure 1).

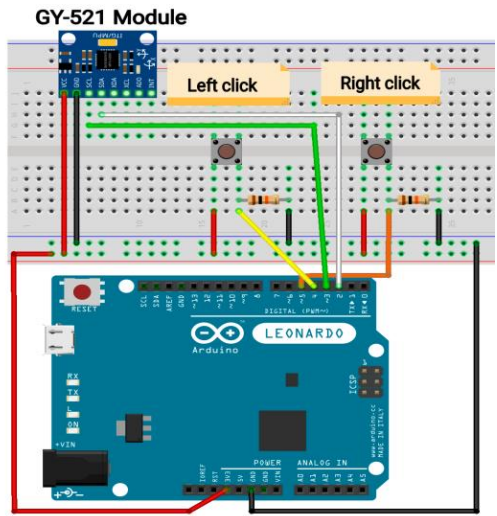

Figure 1. Diagram of the prototype.

### A. Hardware design

The protective casing of the Arduino features: a USB cable used for connection to a computer, a diagnostic indicator system allowing the user to monitor the operational status of the prototype, and 2 female USB ports for accessories A and B [13].

Protective Casing of the Arduino

Figure 2 provides an interior view of the protective casing of the implemented prototype, showcasing the perforated microboard connected to the Arduino with USB ports. These ports facilitate communication between the Arduino Leonardo board, the GY-521 Module, and the foot-operated pedal buttons [14].

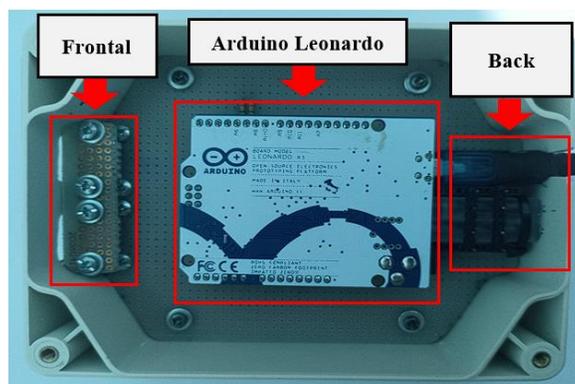

Figure. 2. Interior view of the protective casing of the Arduino-based air mouse prototype.

On the front side of the protective casing, there are USB ports that grant access to connectors for accessories A (device for cursor movement) and B (pedals for right and left-click), as depicted in Figure 3.

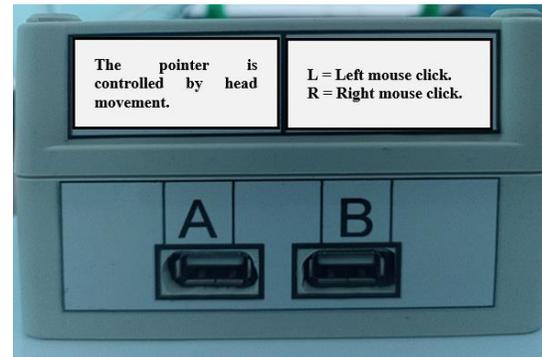

Figure. 3. Front view of the protective casing of the mouse prototype for individuals without upper limbs.

At the back of the protective casing, indicators for the device's operational status and the USB cable used for computer connection are located, as shown in Figure 4.

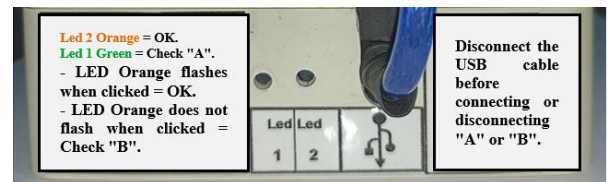

Figure 4. Rear view of the protective casing of the mouse prototype for individuals without upper limbs.

Accessory A

This accessory consists of an elastic band designed to be worn on the head. It includes a small transparent plastic box containing the GY-521 Module [15]. The letter A serves as an identifier for the USB port connection to the prototype. This accessory is illustrated in Figure 5.

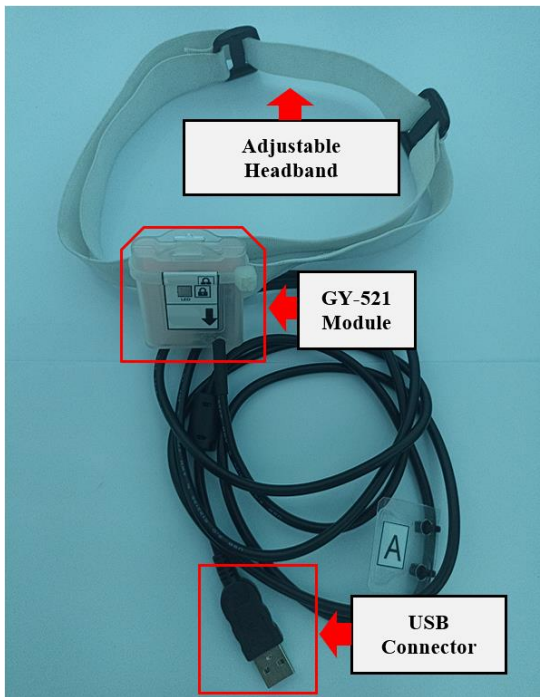

Figure 5. Accessory A that enables pointer movement.

Accessory B

This accessory comprises 2 pedal-shaped buttons [16]. The letters L and R serve as identifiers for left and right-click functions, respectively, while the letter B identifies the USB cable connection to the corresponding port on the prototype. This accessory is depicted in Figure 6.

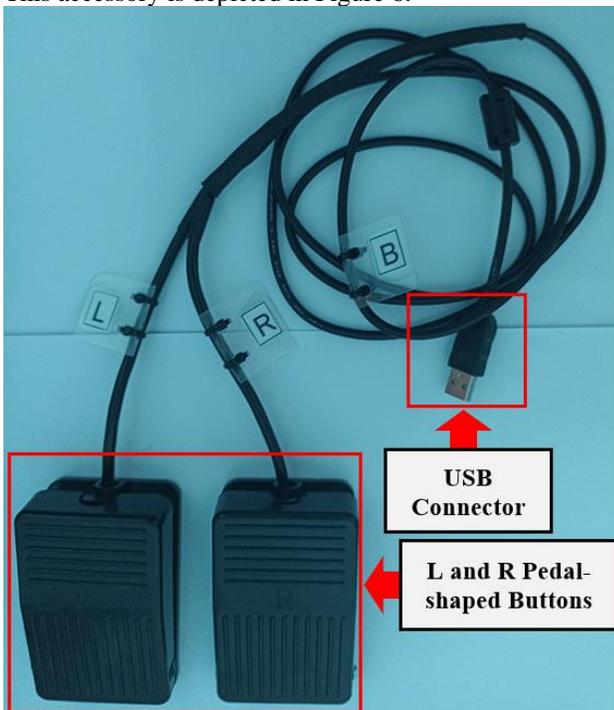

Figure 6. Accessory B, placed on the floor near the user's feet to serve as the prototype's button.

### B. Software design

Programming language: Arduino IDE development platform.

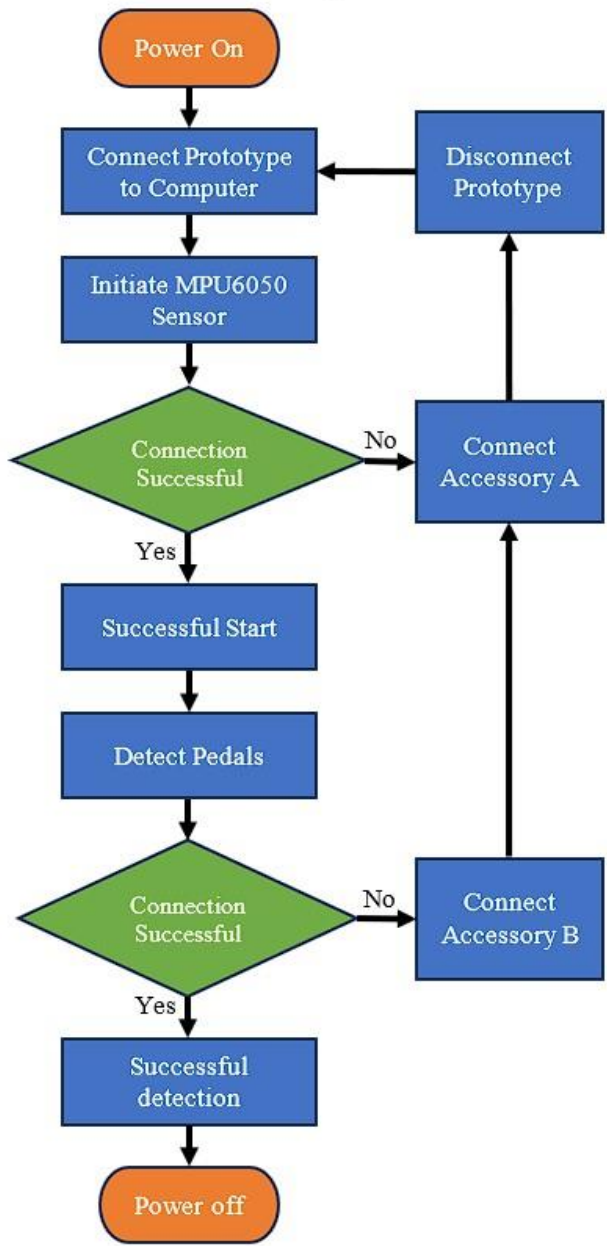

Figure 7. Flowchart depicting the operation of the mouse prototype for individuals without upper limbs.

### C. Prototype development

The USB cable from Accessory A is connected to USB Port A. The elastic band of Accessory A is worn on the head, with the protective box of the GY-521 module positioned to the right, pointing downwards. The two buckles on the elastic band are then used to adjust it to the user's head size, ensuring a secure fit [17].

The USB cable from Accessory B is connected to USB Port B, and the two pedals are placed on the floor near the user's feet. The R pedal is positioned in front of the right foot, and the L pedal is placed in front of the left foot.

The USB cable from the Arduino protective casing is connected to an available USB port on the computer, which will automatically detect the new input device. Diagnostic indicators should be checked, and an orange LED 2 indicates that the prototype and its accessories are functioning correctly [18].

The user utilizes head movements to navigate the computer's graphical interface. The right foot is used to press the R pedal for right-click, and the left foot is used to press the L pedal for left-click, as needed [19].

### D. Design and Performance Testing

The prototype underwent real-world testing, where a user performed everyday actions, such as opening the browser and accessing a webpage, as depicted in Figure 8 and Figure 9. These tests were conducted by connecting the prototype's USB cable to a computer's USB 2.0 port, and it began functioning immediately upon connection without requiring the installation of any additional drivers [20].

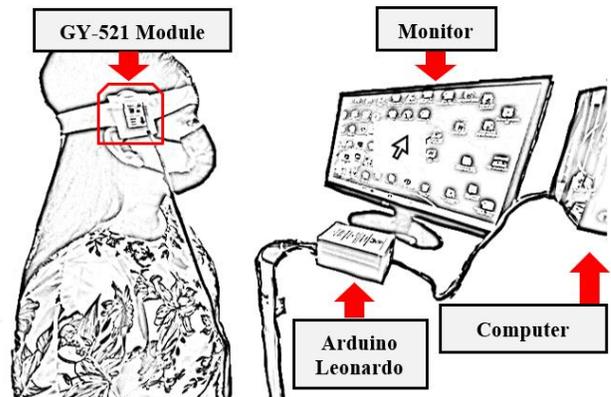

Fig. 8. Functionality test of Accessory A.

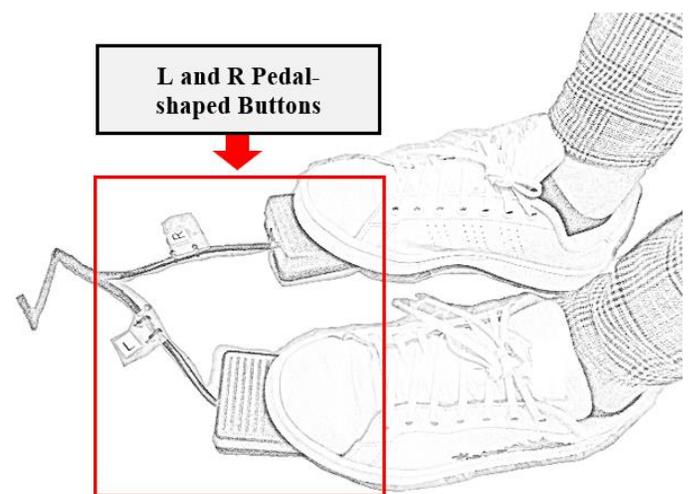

Fig. 9. Functionality test of Accessory B.

## III. RESULTS AND DISCUSSION

To fully meet the set objectives, a comparison was made between the Arduino-based air mouse for individuals without upper limbs and the Genius USB DX-120 mouse. The results are presented in Table I.

Table I. Comparative table of features offered by a common mouse

| Characteristics of a Conventional USB Mouse | Genius USB DX-120 | Mouse for Individuals without Upper Limbs |
|---|---|---|
| Wired USB type A connectivity | Yes | Yes |
| Full cursor control | Yes | Yes |
| Static cursor stability | Yes | No |
| Correct cursor movement (non-erratic) | Yes | Yes |
| Scroll Wheel | Yes | No |
| Compatible with PC and Laptop | Yes | Yes |
| Right and left-click buttons | Yes | Yes |
| Compatible with Windows 10 | Yes | Yes |
| Detects button press when moving | Yes | No |
| Normal delay when detecting button press | Yes | Yes |
| Number of buttons | 3 | 2 |
| Set DPI (Dots Per Inch) value | 1.000 | - |
| Cost | $ 7,00 | $ 52.49 |

After the obtained data analysis, it is evident that the Arduino-based air mouse for individuals without upper limbs provides the user with most of the functions required for computer use. It adequately addresses the needs that may arise in a real-life scenario for a person without upper limbs when using a computer.

## IV. CONCLUSIONS

The results demonstrate that the prototype design is functional and aligns with the user's needs, allowing correct interaction with the computer's graphical interface, including cursor movement, left-click, and right-click. However, it lacks the scroll wheel function. On the flip side, one of its significant advantages is the low manufacturing cost, and the fact that the components required for its assembly are common and easy to acquire.